\documentclass[11pt,a4paper]{article}
\pdfoutput=1
\usepackage[T1]{fontenc}
\usepackage{lmodern,amsmath,amssymb,amsthm,mathtools,bm,microtype}
\usepackage[a4paper,margin=25mm]{geometry}
\usepackage{xcolor,tikz,pgfplots,enumitem}
\pgfplotsset{compat=1.16}
\usepackage[numbers,sort&compress]{natbib}
\usepackage[colorlinks=true,linkcolor=blue!50!black,citecolor=blue!50!black,urlcolor=blue!50!black]{hyperref}
\hypersetup{pdftitle={No smooth real instanton saddles in self-dual Yang-Mills on AdS4},pdfauthor={Bayram Tekin}}
\numberwithin{equation}{section}
\newtheorem{theorem}{Theorem}
\newtheorem{proposition}{Proposition}
\newtheorem{corollary}{Corollary}
\newcommand{\dd}{\mathrm{d}}
\newcommand{\tr}{\operatorname{tr}}
\newcommand{\Vf}{\mathcal V}
\newcommand{\Bd}{\mathcal S}
\newcommand{\Ball}{\mathbb B}
\newcommand{\dens}{\varrho}
\newcommand{\lc}{\tilde\epsilon}
\allowdisplaybreaks[1]

\title{No smooth real instanton saddles in self-dual Yang-Mills on AdS$_4$}
\author{Bayram Tekin\\[5pt]
\small Department of Physics, Bilkent University\\
\small 06800 Ankara, T\"urkiye\\[3pt]
\small\href{mailto:bayram.tekin@bilkent.edu.tr}{bayram.tekin@bilkent.edu.tr}}
\date{November 1, 2026}
\begin{document}
\maketitle
\begin{abstract}
In the self-dual limit of Yang-Mills theory on Euclidean AdS$_4$ studied by
Heuveline, Ruzziconi, Sheta and Strominger, the boundary condition of the
second-order action becomes vacuous, but the boundary equation of the
first-order action is non-trivial. We show that, with this equation, the vacuum is
the only saddle with a real gauge field when both fields are smooth up to the
conformal boundary, for any compact gauge group. In particular, the self-dual
solutions of Sar\i o\u glu and Tekin (2009) are not saddles; we also correct
their charges, which omitted a surface term at the conformal boundary.
Complex saddles and singular limits remain open.
\end{abstract}

\section{Introduction}\label{sec:introduction}

A Yang-Mills instanton is introduced through the self-duality equation $F_{\mu\nu}=\frac12\epsilon_{\mu\nu\rho\sigma}F^{\rho\sigma}$, where $\epsilon_{\mu\nu\rho\sigma}$ is the Levi-Civita tensor. By the Bianchi identity, this equation implies the bulk Yang-Mills equation. However, to be a saddle point of the path integral, the field must also satisfy the boundary equation obtained by varying the action. If the boundary connection is held fixed, its variation vanishes. If we integrate it over, the surface term gives an additional equation. A self-dual field need not satisfy this equation and hence need not be an instanton saddle.

This distinction is crucial for the self-dual Yang-Mills theory on Euclidean AdS$_4$ introduced by Heuveline, Ruzziconi, Sheta, and Strominger (HRSS)~\cite{HRSS2026}. They start from Yang-Mills theory with coupling $g$, a theta angle $\theta$, and Neumann boundary conditions, under which the boundary value of the gauge field is integrated over in the path integral. They split the action into self-dual and anti-self-dual parts with couplings
\begin{equation}
 \frac1{g_\pm^2}=\frac1{g^2}\mp\frac{i\theta}{8\pi^2},
 \label{eq:gpm}
\end{equation}
and take $g_-\to0$ at fixed complex $g_+$. Their first-order formulation adds an auxiliary anti-self-dual antisymmetric tensor $B_{\mu\nu}$ in the adjoint representation of the gauge group. There is a boundary equation that relates $B_{\mu\nu}$ to the field strength. They use this formulation to compute perturbative boundary correlators. HRSS do not study nonperturbative effects, but they suggest that these scale as powers of $e^{-1/g_+^2}$~\cite{HRSS2026}. The natural candidates for such nonperturbative effects are instantons. We therefore ask which classical solutions of the source-free first-order boundary problem have a real gauge field, assuming that both $A$ and $B$ are smooth up to the conformal boundary.

Our starting point was the self-dual solutions found by Sar\i o\u glu and the author in 2009~\cite{ST2009}: we wanted to know whether they become instantons of the HRSS theory and to answer the raised question about nonperturbative effects. As we shall see, they do not solve the boundary equation. They remain self-dual configurations on AdS$_4$, but in the first-order formulation, no auxiliary field satisfies the boundary equation along with them, so they are not saddles. We find that the same obstruction applies to every smooth, non-flat real self-dual connection. In the classical saddle-point sense, the real smooth sector of the theory therefore contains only the vacuum. Whether complex saddles or singular configurations contribute beyond perturbation theory remains open.

The general argument has two steps. First, the bulk equation and the boundary equation for $B$ together imply that the Pontryagin charge $Q$ of the self-dual connection is stationary along every smooth self-dual deformation. Second, Euclidean AdS$_4$ is conformal to a ball, and dilating the ball provides such a deformation: one restricts the field to a smaller concentric ball and stretches it back to unit size. For a real field, the charge then changes at a rate given by the integral of the nonnegative curvature density over the boundary sphere. Stationarity forces the curvature to vanish on the boundary, and unique continuation makes the field flat everywhere. The remaining equations then force $B=0$. No symmetry is assumed; the gauge group is any compact group, and $B$ may be complex. Both the bulk and boundary equations used in the proof follow from the same action.

The boundary equation does not disappear in the self-dual limit. At finite couplings with $g_+^2/g_-^2$ not real, the second-order boundary condition, together with the bulk equations, excludes smooth real non-flat saddles (Sec.~\ref{sec:YMform}). In the second-order formulation, however, this boundary condition carries a factor of the coupling that tends to zero, and in the limit, it is satisfied by every self-dual field; HRSS note that it becomes vacuous. The first-order formulation keeps the information lost in that factor in the independent field $B$, and HRSS keep its boundary equation. Our no-go result uses this surviving boundary equation.

The charges of the $SO(3)$-symmetric solutions from 2009 also need to be corrected. Following Witten~\cite{Witten1977,Tekin2000}, we reduced self-duality to vortex equations on a two-dimensional hyperbolic space and identified the Pontryagin charge with the vortex flux, dropping a surface term. The conformal boundary of AdS$_4$ makes that surface term nonzero in general. As an example, place a BPST instanton~\cite{BPST} of scale $\alpha$ at the center of the unit ball that represents AdS$_4$ (Sec.~\ref{sec:ball}). Normalize the Pontryagin charge so that the complete instanton on $\mathbb R^4$ has charge one. Then the charge inside the ball, $Q$, and the vortex flux, $\Vf$, used in 2009 are
\begin{equation}
 \Vf=\frac1{1+\alpha^2},\qquad
 Q=\frac{1+3\alpha^2}{(1+\alpha^2)^3}.
 \label{eq:introcharges}
\end{equation}
The two agree at $\alpha=1$, so the half-charge solution of 2009 is correct, although the surface term has a nonzero variation there. We derive the surface term from the four-dimensional curvature and check the corrected charge against a direct calculation for a BPST instanton of arbitrary center and size.

The no-go theorem does not use the vortex reduction or the charge correction; even the uncorrected flux varies along the family. We discuss the two results together because the 2009 solutions provide an explicit test of the boundary problem. In both calculations, the field at the conformal boundary is essential.

On $\mathbb R^4$, finite action and regularity lead to the familiar picture of integer charges~\cite{BPST,UhlenbeckRemovable}. On a space with a boundary, smooth self-dual fields with continuously varying charge are natural. Earlier work on curved backgrounds has already shown how much regularity and boundary conditions matter~\cite{BCC79,BCC80,Verbin}, and related self-dual boundary problems in AdS have been studied recently~\cite{SvD2026a,SvD2026b}. Our result is a statement about classical saddles of one specific boundary problem defined in Sec.~\ref{sec:hrss}. It does not evaluate a nonperturbative path integral, and it leaves complex saddles open (Sec.~\ref{sec:beyond}).

\section{The first-order action and its boundary problem}\label{sec:hrss}\label{sec:limit}

Because the boundary value of the gauge field is integrated over, a classical saddle of the HRSS theory must satisfy a boundary equation as well as the bulk equations. We specify the conformal boundary, the regularity of the fields, and our tensor conventions before deriving this equation from the action.

\subsection{The conformal ball and field space}\label{sec:ball}

Euclidean AdS$_4$ of radius $\ell$ is hyperbolic four-space: the open unit ball $\Ball=\{y\in\mathbb R^4:|y|<1\}$ equipped with the complete metric
\begin{equation}
 g_{\mu\nu}=\Omega^2\delta_{\mu\nu},\qquad
 \Omega=\frac{2\ell}{1-r^2},\qquad r=|y|.
 \label{eq:ballmetric}
\end{equation}
We write $g_0$ for the flat metric $\delta_{\mu\nu}$. The sphere $S^3=\{r=1\}$ is not part of this Riemannian manifold. The AdS proper distance from the origin along a radial line is
\begin{equation}
 s(r)=\int_0^r\frac{2\ell\,\dd r'}{1-r'^2}
 =\ell\log\frac{1+r}{1-r}\longrightarrow\infty\qquad(r\to1^-).
 \label{eq:AdSdistance}
\end{equation}
The flat metric $g_0=\Omega^{-2}g$, however, extends smoothly to the closed ball $\bar\Ball=\Ball\cup S^3$. Adding $S^3$ describes limiting directions and field data at AdS infinity; it does not add a physical wall at a finite AdS distance. Surface terms can equivalently be evaluated on cutoff spheres $S_R=\{r=R<1\}$ and then taken to the limit $R\to1^-$. We call a field \emph{smooth} if, in a suitable gauge, its components $A_\mu(y)$ in Cartesian coordinates $y^\mu$ (and likewise $B_{\mu\nu}(y)$ for the auxiliary field of Sec.~\ref{sec:action}) extend to smooth functions on $\bar\Ball$. This is a regularity assumption at the conformal boundary, which is stronger than finite action. The lower-index components $A_\mu$ and $B_{\mu\nu}$ are not rescaled by powers of $\Omega$; in this sense, they have conformal weight zero.

\paragraph{Gauge fields.}
The explicit solutions use $SU(2)$ with anti-Hermitian generators $T_a=-i\sigma_a/2$, so that $[T_a,T_b]=\epsilon_{abc}T_c$ and $\tr(T_aT_b)=-\delta_{ab}/2$. The general results use a compact Lie algebra with a positive invariant inner product; in an orthonormal basis $T_a$ the structure constants $f_{abc}$, defined by $[T_a,T_b]=f_{abc}T_c$, are real and totally antisymmetric ($f_{abc}=\epsilon_{abc}$ for $SU(2)$). A connection is real when its components $A^a_\mu$ in this basis are real. We write $A_\mu=A^a_\mu T_a$ and
\begin{equation}
 F_{\mu\nu}=\partial_\mu A_\nu-\partial_\nu A_\mu+[A_\mu,A_\nu],\qquad
 F^a_{\mu\nu}=\partial_\mu A^a_\nu-\partial_\nu A^a_\mu+f_{abc}A^b_\mu A^c_\nu .
 \label{eq:conventions}
\end{equation}
Here $A_\mu$ is a covariant vector and $F_{\mu\nu}$ an antisymmetric covariant tensor.  Adjoint indices are raised and lowered with $\delta_{ab}$.

\paragraph{Metric, duality and pairing.}
Unless the flat frame is specified, spacetime indices are raised with $g^{\mu\nu}$ and the volume element is $\sqrt g\,\dd^4y$. The Cartesian coordinates are positively oriented, and the Levi-Civita tensor is $\epsilon_{\mu\nu\rho\sigma}=\sqrt g\,\lc_{\mu\nu\rho\sigma}$, where $\lc_{\mu\nu\rho\sigma}$ is the totally antisymmetric Levi-Civita symbol, with $\lc_{1234}=1$ and entries $0,\pm1$ in every coordinate system; thus $\epsilon_{1234}=1$ in the flat frame and $\Omega^4$ in the AdS frame. The dual of an antisymmetric tensor $X_{\mu\nu}$ is
\begin{equation}
 (\star X)_{\mu\nu}=\tfrac12\epsilon_{\mu\nu\rho\sigma}X^{\rho\sigma}.
 \label{eq:Hodgecurved}
\end{equation}
In four Euclidean dimensions $\star\star X=X$, so every antisymmetric tensor splits into a self-dual part ($\star X=X$) and an anti-self-dual part ($\star X=-X$). For adjoint-valued antisymmetric tensors, we use the bilinear pairing
\begin{equation}
 X\cdot Y=X^a_{\mu\nu}Y^{a\mu\nu},
 \qquad X\cdot\star Y=\star X\cdot Y .
 \label{eq:pairing}
\end{equation}
The second identity holds because $\epsilon_{\mu\nu\rho\sigma}$ is symmetric under the exchange of the index pairs $(\mu\nu)\leftrightarrow(\rho\sigma)$. It implies the fact used repeatedly below: if $X$ is self-dual and $Y$ anti-self-dual, then $X\cdot Y=\star X\cdot Y=X\cdot\star Y=-X\cdot Y$, so $X\cdot Y=0$. No complex conjugation enters the pairing, so this holds for complex fields as well. Conjugation will be inserted explicitly when a positive norm for a complex $B$ is needed.

\paragraph{Two covariant derivatives.}
Two connections act on our fields: the Levi-Civita connection of $g$ on spacetime indices and the gauge connection on the adjoint index. We write $\nabla_\mu$ for the first alone and
\begin{equation}
 D_\mu X_{\nu_1\cdots\nu_p}
 =\nabla_\mu X_{\nu_1\cdots\nu_p}+[A_\mu,X_{\nu_1\cdots\nu_p}]
 =\partial_\mu X_{\nu_1\cdots\nu_p}
   -\sum_{k=1}^p\Gamma^\lambda_{\mu\nu_k}
       X_{\nu_1\cdots\lambda\cdots\nu_p}
   +[A_\mu,X_{\nu_1\cdots\nu_p}]
 \label{eq:Dcomponents}
\end{equation}
for both together on an adjoint-valued tensor, with the usual $+\Gamma$ terms for upper indices. On a gauge-invariant (colorless) quantity, $D_\mu=\nabla_\mu$; on a spacetime scalar, $\nabla_\mu=\partial_\mu$; in flat Cartesian coordinates, $\nabla_\mu=\partial_\mu$ on every tensor. Three consequences are used throughout.
\begin{enumerate}[label=(\roman*),nosep]
\item \emph{Antisymmetrized derivatives.} Because the Levi-Civita connection is torsion-free, $\Gamma^\lambda_{\mu\nu}=\Gamma^\lambda_{\nu\mu}$, and the Christoffel terms cancel in antisymmetric combinations. For example, $D_\mu a_\nu-D_\nu a_\mu=\partial_\mu a_\nu-\partial_\nu a_\mu+[A_\mu,a_\nu]-[A_\nu,a_\mu]$ for an adjoint-valued vector $a_\mu$. The Bianchi identity reads
\begin{equation}
 D_\mu F_{\nu\rho}+D_\nu F_{\rho\mu}+D_\rho F_{\mu\nu}=0
 \quad\Longleftrightarrow\quad
 D_\mu(\star F)^{\mu\nu}=0,
 \label{eq:covariant}
\end{equation}
the two statements being related by contraction with $\epsilon^{\sigma\mu\nu\rho}$, which is covariantly constant.
\item \emph{Divergence of an antisymmetric tensor.} For $X^{\mu\nu}=-X^{\nu\mu}$ the term $\Gamma^\nu_{\mu\lambda}X^{\mu\lambda}$ vanishes, so
\begin{equation}
 D_\mu X^{\mu\nu}=\frac1{\sqrt g}\,\partial_\mu\big(\sqrt g\,X^{\mu\nu}\big)+[A_\mu,X^{\mu\nu}] .
 \label{eq:Bdivergence}
\end{equation}
\item \emph{Covariant Leibniz rule.} Invariance of the inner product means $f_{abc}$ is totally antisymmetric, and then the gauge terms cancel in
\begin{equation}
 \nabla_\mu\big(X^{a\mu\nu}a^a_\nu\big)=(D_\mu X^{\mu\nu})^a a^a_\nu+X^{a\mu\nu}(D_\mu a_\nu)^a .
 \label{eq:leibniz}
\end{equation}
The left side is the divergence of the vector $V^\mu=X^{a\mu\nu}a^a_\nu$, equal to $\frac1{\sqrt g}\partial_\mu(\sqrt g\,V^\mu)$, so Gauss's theorem applies to it.
\end{enumerate}

\paragraph{The charge.}
The Pontryagin charge is
\begin{equation}
 Q[A]=\frac1{32\pi^2}\int_\Ball\dd^4y\,\sqrt g\,F^a_{\mu\nu}(\star F)^{a\mu\nu}
 =\frac1{64\pi^2}\int_\Ball\dd^4y\,\lc_{\mu\nu\rho\sigma}\,F^a_{\mu\nu}F^a_{\rho\sigma},
 \label{eq:Q}
\end{equation}
where the second expression shows that $Q$ does not depend on the metric. It is normalized so that a BPST instanton on all of $\mathbb R^4$ has $Q=1$. For a self-dual field, it may be written as $Q=\int_\Ball\dens\,\dd^4y$ with the flat-space density
\begin{equation}
 \dens=\frac1{32\pi^2}\sum_{a,\mu,\nu}\big(F^a_{\mu\nu}\big)^2\qquad(F=\star F),
 \label{eq:flatdensity}
\end{equation}
which is nonnegative for a real connection. The integrand of \eqref{eq:Q} is a total derivative, with the Chern--Simons current
\begin{equation}
 K^\mu=\lc_{\mu\nu\rho\sigma}\Big(A^a_\nu\partial_\rho A^a_\sigma+\tfrac13f_{abc}A^a_\nu A^b_\rho A^c_\sigma\Big),
 \qquad
 \tfrac12\lc_{\mu\nu\rho\sigma}F^a_{\mu\nu}F^a_{\rho\sigma}=2\,\partial_\mu K^\mu .
 \label{eq:Kcurrent}
\end{equation}
Gauss's theorem in the flat frame therefore gives, in any gauge smooth on $\bar\Ball$,
\begin{equation}
 Q=\frac1{16\pi^2}\oint_{S^3}\dd\Sigma\;n_\mu K^\mu ,
 \label{eq:CS}
\end{equation}
With $n^\mu=y^\mu$ as the outward unit normal and $\dd\Sigma$ as the area element of the unit sphere. Contracting $\lc_{\mu\nu\rho\sigma}$ with $n_\mu$ leaves only components of $A$ tangent to $S^3$ and their tangential derivatives, so $Q$ is determined by the boundary value of the connection. $K^\mu$ is not gauge invariant, but under a gauge transformation smooth on $\bar\Ball$ its flux changes by the winding number of the transformation restricted to $S^3$, which is zero. When the boundary connection is free and not flat, $Q$ need not be an integer.

\subsection{From Yang-Mills with a theta term to the first-order action}\label{sec:action}

Let us start with the usual Euclidean Yang-Mills action and a theta term, with the sign fixed by \eqref{eq:gpm}:
\begin{align}
 S_{\rm YM,\theta}[A]
 &=\frac1{4g^2}\int_\Ball\dd^4y\sqrt g\,F^a_{\mu\nu}F^{a\mu\nu}-i\theta Q[A]\nonumber\\
 &=\int_\Ball\dd^4y\sqrt g\left[
 \frac1{4g^2}F\cdot F
 -\frac{i\theta}{32\pi^2}F\cdot\star F\right].
 \label{eq:YMtheta}
\end{align}
Write $F^\pm=\frac12(F\pm\star F)$, so that $\star F^\pm=\pm F^\pm$. By \eqref{eq:pairing}, $F^+\cdot F^-=0$, and therefore
\begin{equation}
 F\cdot F=F^+\cdot F^++F^-\cdot F^-,\qquad
 F\cdot\star F=F^+\cdot F^+-F^-\cdot F^-.
 \label{eq:chiralidentities}
\end{equation}
Substitution into \eqref{eq:YMtheta} gives the two-coupling action of HRSS :
\begin{equation}
 S[A]=\frac1{4g_+^2}\int_\Ball\dd^4y\sqrt g\,F^{+a}_{\mu\nu}F^{+a\mu\nu}
 +\frac1{4g_-^2}\int_\Ball\dd^4y\sqrt g\,F^{-a}_{\mu\nu}F^{-a\mu\nu},
 \label{eq:S2}
\end{equation}
with $g_\pm^{-2}=g^{-2}\mp i\theta/(8\pi^2)$ as in \eqref{eq:gpm}. The self-dual limit is $g_-\to0$ at fixed $g_+$. Equivalently,
\begin{equation}
 \frac1{g^2}=\frac12\left(\frac1{g_+^2}+\frac1{g_-^2}\right),\qquad
 \theta=4\pi^2i\left(\frac1{g_+^2}-\frac1{g_-^2}\right).
 \label{eq:thetacontinuation}
\end{equation}
Thus, both $g^{-2}$ and $\theta$ have divergent parts in this limit, while their combination $g^{-2}-i\theta/(8\pi^2)=g_+^{-2}$ stays fixed. This is an analytic continuation in coupling space, not a limit at a fixed real theta angle. For real $g$ and $\theta$, $g_+^{-2}$ and $g_-^{-2}$ are complex conjugates and cannot be taken to these independent limits.

To introduce the auxiliary field, first rewrite \eqref{eq:S2} as
\begin{equation}
 S[A]=\int_\Ball\dd^4y\sqrt g\left[
 \frac1{4\varepsilon}F^-\cdot F^-
 +\frac1{4g_+^2}F\cdot\star F\right],\qquad
 \frac1\varepsilon=\frac1{g_+^2}+\frac1{g_-^2}.
 \label{eq:actionrewrite}
\end{equation}
The auxiliary field is an adjoint-valued antisymmetric tensor $B_{\mu\nu}$ in the same anti-self-dual subspace as $F^-$: $\star B=-B$. This choice is natural because a self-dual component would have zero pairing with $F^-$ and would not couple to it. It is part of the first-order formulation, not an additional equation imposed on $F$. For anti-self-dual $B$ and $\varepsilon\ne0$, completing the square gives
\begin{equation}
 B\cdot F^--\varepsilon B\cdot B
 =\frac1{4\varepsilon}F^-\cdot F^-
 -\varepsilon\left(B-\frac{F^-}{2\varepsilon}\right)
                  \cdot\left(B-\frac{F^-}{2\varepsilon}\right).
 \label{eq:completesquare}
\end{equation}
The last term vanishes at its stationary point $B=F^-/(2\varepsilon)$. Formally, integrating over the shifted field along a suitable Gaussian contour contributes only a field-independent normalization. Replacing the first term in \eqref{eq:actionrewrite} by the left-hand side therefore gives the HRSS first-order action (their Eq.~(3.1)), of the type introduced in Refs.~\cite{ChalmersSiegel,CattaneoBF}:
\begin{equation}
 S_\varepsilon[A,B]=\int_\Ball\dd^4y\sqrt g\left[
 B^a_{\mu\nu}F^{-a\mu\nu}-\varepsilon B^a_{\mu\nu}B^{a\mu\nu}
 +\frac1{4g_+^2}F^a_{\mu\nu}(\star F)^{a\mu\nu}\right].
 \label{eq:SBF}
\end{equation}
Variation of $B$ within the anti-self-dual subspace gives
\begin{equation}
 F^-=2\varepsilon B,\qquad
 \varepsilon=\frac{g_+^2g_-^2}{g_+^2+g_-^2},\qquad
 (2\varepsilon)^{-1}=\tfrac12(g_+^{-2}+g_-^{-2})=g^{-2}.
 \label{eq:epsilon}
\end{equation}
Eliminating $B$ at finite couplings recovers \eqref{eq:S2}; equivalently $B=F^-/g^2$. As $g_-\to0$, $\varepsilon\to0$, and the independent field $B$ becomes a Lagrange multiplier imposing $F^-=0$. We denote the limiting action by $S_0$. By ``BF formulation'' we mean this chiral first-order action, including its nonzero Pontryagin term.

\paragraph{Conformal weights.}
The covariant components $F_{\mu\nu}$ and $B_{\mu\nu}$ are held fixed under $g=\Omega^2g_0$, consistently with $B=F^-/g^2$ at finite couplings. For either tensor $X_{\mu\nu}$,
\begin{equation}
 \sqrt g=\Omega^4\sqrt{g_0},\qquad
 X^{\mu\nu}_{(g)}=\Omega^{-4}X^{\mu\nu}_{(0)},\qquad
 (\star_gX)_{\mu\nu}=(\star_0X)_{\mu\nu},
 \label{eq:conformalweights}
\end{equation}
where the subscripts indicate the metric used to raise indices or to define the dual. The last identity follows from \eqref{eq:Hodgecurved}: $\epsilon_{\mu\nu\rho\sigma}$ contributes $\Omega^4$ and the two raised indices $\Omega^{-4}$; it is special to antisymmetric tensors with two indices in four dimensions, and it preserves self-duality and anti-self-duality. In every term of \eqref{eq:SBF} the two raised indices contribute $\Omega^{-4}$, which cancels the factor $\Omega^4$ in $\sqrt g$. In particular,
\begin{equation}
 \sqrt g\,\varepsilon B^a_{\mu\nu}B^{a\mu\nu}_{(g)}
 =\sqrt{g_0}\,\varepsilon B^a_{\mu\nu}B^{a\mu\nu}_{(0)}.
 \label{eq:Bquadraticconformal}
\end{equation}
All three terms of the action are therefore conformally invariant, including the quadratic $B$ term, for any fixed spacetime-independent couplings. Working on the flat conformal ball is valid already at finite $\varepsilon$; the self-dual limit removes the quadratic term but is not needed for conformal invariance.

\subsection{Field equations and the boundary equation}\label{sec:BFform}

To derive the equations in the AdS frame, first restrict the action to $M_R=\{r\le R<1\}$ with the AdS metric $g$, and call the restricted action $S_{\varepsilon,R}$. Let $n^{(g)}_\mu$ be the outward unit normal covector of $S_R=\partial M_R$ and $\dd\Sigma_g$ its induced area element. The cutoff is used only to evaluate the variation; the physical boundary equation will be imposed on its limit at conformal infinity, not on every cutoff sphere.

Vary $A_\mu$ by the adjoint-valued vector $a_\mu=\delta A_\mu$, keeping the metric fixed. From \eqref{eq:conventions},
\begin{equation}
 \delta F_{\mu\nu}=D_\mu a_\nu-D_\nu a_\mu,\qquad
 D_\mu a_\nu=\partial_\mu a_\nu-\Gamma^\lambda_{\mu\nu}a_\lambda+[A_\mu,a_\nu].
 \label{eq:deltaF}
\end{equation}
The Christoffel terms cancel in the antisymmetric difference, but they must be kept when integrating by parts, which is done with the full $D_\mu$ through \eqref{eq:leibniz}. Since $B$ is anti-self-dual, its pairing with the self-dual part of $\delta F$ vanishes, so $B\cdot\delta F^-=B\cdot\delta F$; antisymmetry of $B^{\mu\nu}$ then gives
\begin{equation}
 \delta_A\int_{M_R}\dd^4y\sqrt g\,B\cdot F^-
 =2\int_{M_R}\dd^4y\sqrt g\,B^{a\mu\nu}(D_\mu a_\nu)^a .
 \label{eq:BFvariation}
\end{equation}
By \eqref{eq:leibniz}, $B^{a\mu\nu}(D_\mu a_\nu)^a=\nabla_\mu(B^{a\mu\nu}a^a_\nu)-(D_\mu B^{\mu\nu})^aa^a_\nu$, where $D_\mu B^{\mu\nu}$ is given by \eqref{eq:Bdivergence}. For the Pontryagin term, $F\cdot\star\delta F=\delta F\cdot\star F$ by \eqref{eq:pairing}, so
\begin{equation}
 \delta(F\cdot\star F)=2\,(\star F)\cdot\delta F
 =4(\star F)^{a\mu\nu}(D_\mu a_\nu)^a,
 \label{eq:topvariation}
\end{equation}
and the same Leibniz rule turns this into a total divergence, because $D_\mu(\star F)^{\mu\nu}=0$ by the Bianchi identity \eqref{eq:covariant}. Gauss's theorem then gives the full variation
\begin{align}
 \delta S_{\varepsilon,R}={}&\int_{M_R}\dd^4y\sqrt g\left[
 (F^{-a\mu\nu}-2\varepsilon B^{a\mu\nu})\delta B^a_{\mu\nu}
 -2(D_\mu B^{\mu\nu})^a a^a_\nu\right]\nonumber\\
 &+2\int_{S_R}\dd\Sigma_g\,n^{(g)}_\mu
 \left[B^{a\mu\nu}+\frac1{2g_+^2}(\star F)^{a\mu\nu}\right]a^a_\nu .
 \label{eq:fullvariation}
\end{align}
Here $\delta B$ is anti-self-dual, as is its coefficient. Interior variations give, in the limit $\varepsilon\to0$,
\begin{equation}
 F^-=0,\qquad D_\mu B^{\mu\nu}=0,\qquad\star B=-B.
 \label{eq:bulkequations}
\end{equation}

To take the boundary limit, use on the same sphere $S_R$ the flat unit normal $n_\mu$ and flat area element $\dd\Sigma$. For any antisymmetric tensor $X_{\mu\nu}$,
\begin{equation}
 \dd\Sigma_g=\Omega^3\dd\Sigma,\qquad
 n^{(g)}_\mu=\Omega\,n_\mu,\qquad
 \dd\Sigma_g\,n^{(g)}_\mu X^{\mu\nu}_{(g)}
 =\dd\Sigma\,n_\mu X^{\mu\nu}_{(0)}.
 \label{eq:boundaryweights}
\end{equation}
Consequently the surface term has a finite limit as $R\to1^-$, which smoothness on $\bar\Ball$ guarantees, and it is given by the flat-frame expression on $S^3$. \emph{From here on we work in the flat frame:} indices are raised with $\delta^{\mu\nu}$, the volume element is $\dd^4y$, $\star=\star_0$, and $n_\mu$, $\dd\Sigma$ are the flat normal and area element. In Cartesian coordinates $D_\mu X=\partial_\mu X+[A_\mu,X]$ on every adjoint-valued tensor, since all Christoffel symbols vanish; whenever curvilinear coordinates are used later, $D_\mu$ again includes the Levi-Civita connection of the metric in use.

Since the boundary connection is free to vary, the tangential components of $a_\nu$ on $S^3$ are arbitrary, and their coefficient in the limiting surface term must vanish:
\begin{equation}
 n_\mu\left(B^{\mu\nu}+\frac1{2g_+^2}(\star F)^{\mu\nu}\right)\Big|_{S^3}=0 .
 \label{eq:NBF}
\end{equation}
This is Eq.~(3.4) of Ref.~\cite{HRSS2026}, now read in the flat conformal frame. Its normal component, $n_\mu n_\nu(\ldots)^{\mu\nu}$, vanishes identically by antisymmetry, so \eqref{eq:NBF} is three equations per generator. Using the AdS unit normal alone, without the area and index weights in \eqref{eq:boundaryweights}, would introduce a factor $\Omega^{-1}\to0$ and would incorrectly make the boundary equation look empty.

\subsection{What survives the self-dual limit}\label{sec:YMform}\label{sec:vacuous}

At finite couplings, the field equation \eqref{eq:epsilon} turns the boundary coefficient into
\begin{equation}
 2B+g_+^{-2}\star F=g_+^{-2}F^++g_-^{-2}F^- .
 \label{eq:finiteboundary}
\end{equation}
Multiplying the boundary equation by $g_+^2g_-^2$ gives the boundary condition of the second-order action, Eq.~(2.7) of Ref.~\cite{HRSS2026}:
\begin{equation}
 n^\mu\bigl(g_-^2F^+_{\mu\nu}+g_+^2F^-_{\mu\nu}\bigr)\Big|_{S^3}=0 .
 \label{eq:N2}
\end{equation}

A local picture helps. At a point of $S^3$ choose Cartesian axes with $n=\partial_1$, and let $i,j,k$ run over the tangential directions $2,3,4$. Then $n^\mu X_{\mu\nu}$ consists of the ``electric'' components $X_{1i}$, while
\begin{equation}
 n^\mu(\star X)_{\mu i}=(\star X)_{1i}=\tfrac12\lc_{1ijk}\,X_{jk}
 \label{eq:localdual}
\end{equation}
involves only the tangential (``magnetic'') components $X_{jk}$. For a self-dual tensor, the electric and magnetic components therefore determine each other: $X_{1i}=\frac12\lc_{1ijk}X_{jk}$; for an anti-self-dual tensor, the same holds with a minus sign.

At finite couplings with $g_+^2/g_-^2$ not real, \eqref{eq:N2} together with the bulk equations excludes smooth, real, non-flat saddles. The bulk equation of \eqref{eq:S2} is $D_\mu(g_+^{-2}F^{+\mu\nu}+g_-^{-2}F^{-\mu\nu})=g^{-2}D_\mu F^{\mu\nu}=0$, the ordinary Yang-Mills equation, because $D_\mu(\star F)^{\mu\nu}=0$. Write the boundary condition as $g_+^{-2}n^\mu F^+_{\mu\nu}+g_-^{-2}n^\mu F^-_{\mu\nu}=0$. For real $A$, both $n^\mu F^\pm_{\mu\nu}$ are real, so if $g_+^2/g_-^2$ is not real, they vanish separately. This happens for finite real nonzero $g$ and real $\theta\ne0$, since $g_-^{-2}=\overline{g_+^{-2}}$ and $g_+^{-2}$ has nonzero real and imaginary parts; it also happens along generic paths into the self-dual limit. Then $n^\mu F_{\mu\nu}=0$ and $n^\mu(\star F)_{\mu\nu}=0$. By \eqref{eq:localdual}, the second condition states that the tangential components $F_{jk}$ vanish, and the first states that the normal components $F_{1i}$ vanish; thus, the full curvature vanishes on $S^3$. Proposition~\ref{prop:uc}, in its Yang-Mills form, makes $A$ flat. At these couplings, the vacuum is the only smooth, real solution of the source-free Neumann problem, whether or not it is self-dual.

At $g_-=0$, however, \eqref{eq:N2} reads $n^\mu F^-_{\mu\nu}=0$ and is satisfied by every self-dual field; HRSS note that it becomes vacuous (their footnote~9). Multiplication by $g_-^2$ has discarded the finite independent field $B$ in \eqref{eq:finiteboundary}. The limiting first-order action retains its boundary equation \eqref{eq:NBF}, which must still be imposed. This is the boundary problem studied here. Whether the two path-integral formulations agree beyond perturbation theory on particular integration cycles is a separate question.

For self-dual $F$ we have $\star F=F$, so the complete limiting problem is
\begin{equation}
 \boxed{\;F=\star F,\qquad\star B=-B,\qquad D_\mu B^{\mu\nu}=0,
 \qquad n_\mu\big(B^{\mu\nu}+c\,F^{\mu\nu}\big)\Big|_{S^3}=0,\qquad c=\frac1{2g_+^2}\ne0.\;}
 \label{eq:critical}
\end{equation}
In the local axes of \eqref{eq:localdual} the boundary equation reads $B_{1i}=-cF_{1i}$; since $B$ is anti-self-dual and $F$ self-dual, this is equivalent to $B_{jk}=c\,F_{jk}$ for the tangential components. The tangential part of $B$ on the boundary must equal $c$ times the curvature of the boundary connection. Both fields are smooth on $\bar\Ball$, there are no boundary sources or added boundary terms, and the boundary connection is free to vary. The choice $B=0$ would force the boundary curvature to vanish, but that alone does not rule out other choices. The next section rules them all out for real non-flat $A$.

\section{No smooth real non-flat saddles}\label{sec:main}

\subsection{A necessary stationarity condition}

The role of $B$ is that of a Lagrange multiplier. In finite dimensions, stationary points of $S(x)$ subject to a regular constraint $G(x)=0$ obey $\vec\nabla S=\mu\vec\nabla G$. For the multiplier action $S(x)-\mu G(x)$, varying $\mu$ gives $G=0$, while varying $x$ gives $\vec\nabla S=\mu\vec\nabla G$. In the self-dual theory, the constraint is $F^-=0$, and $B$ is its multiplier. On $F^-=0$ we have $F\cdot\star F=F\cdot F$, and \eqref{eq:Q} gives
\begin{equation}
 S_0[A,B]=\frac1{4g_+^2}\int_\Ball\dd^4y\,F\cdot\star F=\frac{8\pi^2}{g_+^2}\,Q[A]\qquad(F^-=0).
 \label{eq:onshell}
\end{equation}
Along a family of self-dual connections, the constraint term $B\cdot F^-$ vanishes identically for any fixed $B$. A stationary pair $(A,B)$ must therefore make $Q$ stationary along every such family, and no choice of $B$ can compensate for a change in the charge. The following theorem derives this directly from \eqref{eq:critical}, including the boundary.

\begin{theorem}[Stationarity]\label{thm:stationary}
Let $M$ be a compact oriented four-dimensional Riemannian manifold with boundary $\partial M$, outward unit normal $n_\mu$ and boundary area element $\dd\Sigma$, and define $Q$ on $M$ by \eqref{eq:Q}. Let $A_\mu(s)$ be a family of smooth self-dual connections on $M$, differentiable in $s$, with $A=A(0)$ and $a_\mu=\partial_sA_\mu|_{s=0}$. Suppose there is a smooth, possibly complex, anti-self-dual adjoint tensor $B_{\mu\nu}$ with
\begin{equation*}
 D_\mu B^{\mu\nu}=0\ \text{in }M,\qquad
 n_\mu\big(B^{\mu\nu}+c\,(\star F)^{\mu\nu}\big)=0\ \text{on }\partial M,\qquad c\ne0 .
\end{equation*}
Then $\dd Q[A(s)]/\dd s=0$ at $s=0$.
\end{theorem}

\begin{proof}
\emph{Step 1.} Differentiating $F_{\mu\nu}(s)=(\star F(s))_{\mu\nu}$ at $s=0$ shows that $\delta F_{\mu\nu}=D_\mu a_\nu-D_\nu a_\mu$ is self-dual.

\emph{Step 2: a conserved current.} Define $J^\mu=B^{a\mu\nu}a^a_\nu$. By the Leibniz rule \eqref{eq:leibniz},
\begin{equation}
 \nabla_\mu J^\mu=(D_\mu B^{\mu\nu})^aa^a_\nu+B^{a\mu\nu}(D_\mu a_\nu)^a
 =0+\tfrac12\,B\cdot\delta F=0 .
 \label{eq:closedpair}
\end{equation}
The first term vanishes by the bulk equation for $B$. In the second, antisymmetry of $B^{\mu\nu}$ gives $B^{a\mu\nu}(D_\mu a_\nu)^a=\frac12B\cdot\delta F$, which vanishes because $B$ is anti-self-dual and $\delta F$ is self-dual, by \eqref{eq:pairing}; no reality is needed.

\emph{Step 3: Gauss's theorem and the boundary equation.} Integrating \eqref{eq:closedpair} over $M$,
\begin{equation}
 0=\oint_{\partial M}\dd\Sigma\,n_\mu B^{a\mu\nu}a^a_\nu
 =-c\oint_{\partial M}\dd\Sigma\,n_\mu(\star F)^{a\mu\nu}a^a_\nu .
 \label{eq:pair}
\end{equation}

\emph{Step 4: the variation of the charge.} By \eqref{eq:Q} and \eqref{eq:topvariation},
\begin{align}
 \frac{\dd Q}{\dd s}\Big|_0
 &=\frac1{8\pi^2}\int_M\dd^4y\sqrt g\,(\star F)^{a\mu\nu}(D_\mu a_\nu)^a
 =\frac1{8\pi^2}\int_M\dd^4y\sqrt g\,\nabla_\mu\big((\star F)^{a\mu\nu}a^a_\nu\big)\nonumber\\
 &=\frac1{8\pi^2}\oint_{\partial M}\dd\Sigma\,n_\mu(\star F)^{a\mu\nu}a^a_\nu ,
 \label{eq:dQds}
\end{align}
where the middle step uses \eqref{eq:leibniz} and the Bianchi identity $D_\mu(\star F)^{\mu\nu}=0$.

\emph{Step 5.} Since $c\ne0$, \eqref{eq:pair} sets the boundary integral in \eqref{eq:dQds} to zero. Hence $\dd Q/\dd s=0$.
\end{proof}

The stationarity condition holds for every smooth auxiliary field that completes $A$, including complex $B$, and the proof also allows complex $A$. Equation \eqref{eq:dQds} shows that the first variation of $Q$ depends only on the variation of the boundary connection.

\subsection{Dilation and positivity}\label{sec:dilation}

For $0<\lambda\le1$ define the dilated field
\begin{equation}
 (A_\lambda)_\mu(y)=\lambda A_\mu(\lambda y),\qquad
 (F_\lambda)_{\mu\nu}(y)=\lambda^2F_{\mu\nu}(\lambda y).
 \label{eq:dilatation}
\end{equation}
That is, $A_\lambda$ is $A$ restricted to the ball of radius $\lambda$ and stretched back to unit size. It is self-dual whenever $A$ is, since in the flat frame $(\star F_\lambda)_{\mu\nu}(y)=\frac12\lc_{\mu\nu\rho\sigma}\lambda^2F_{\rho\sigma}(\lambda y)=\lambda^2(\star F)_{\mu\nu}(\lambda y)$. The family is smooth on the closed ball, equals $A$ at $\lambda=1$, and tends smoothly to the zero connection as $\lambda\to0$. It changes the boundary data, which the Neumann problem allows; it is not an isometry of AdS$_4$. With $\dd\Omega_3$ the volume element of the unit sphere and $\omega$ a point on it, the substitution $x=\lambda y$ in \eqref{eq:flatdensity} gives
\begin{align}
 Q[A_\lambda]
 &=\int_{|y|<1}\lambda^4\dens(\lambda y)\,\dd^4y
 =\int_{|x|<\lambda}\dens(x)\,\dd^4x,\nonumber\\
 \frac{\dd Q[A_\lambda]}{\dd\lambda}
 &=\lambda^3\int_{S^3}\dens(\lambda\omega)\,\dd\Omega_3 .
 \label{eq:dQdlambda}
\end{align}
For a real field, the derivative is nonnegative. Every smooth self-dual field is thus connected to the vacuum by a path of self-dual fields along which the charge grows continuously from zero. When the boundary data are free, the charge does not label disconnected sectors.

We also need a uniqueness result at the boundary.

\begin{proposition}\label{prop:uc}
Let $A$ be a real connection, smooth on $\bar\Ball$, whose field strength $F_{\mu\nu}$ vanishes at every point of $S^3$ (all components). If $A$ is self-dual, or more generally satisfies the Yang-Mills equations, then $F_{\mu\nu}=0$ on all of $\bar\Ball$.
\end{proposition}

A self-dual connection satisfies the Yang-Mills equations by the Bianchi
identity, so it suffices to prove the Yang-Mills statement, which is also
used in the finite-coupling discussion above.

\begin{proof}
Since the curvature vanishes on $S^3$, the boundary connection is flat, and
since $S^3$ is simply connected it is pure gauge. Choose a gauge in a collar
of $S^3$ with $A_r=0$ and $A_i|_{r=1}=0$. Since the full curvature vanishes
on $S^3$, also $\partial_r A_i|_{r=1}=F_{ri}|_{r=1}=0$. In this gauge the
angular Yang-Mills equations in the flat ball metric read
\begin{equation}
 \partial_r^2 A_i+\frac1r\partial_r A_i
 +\frac1{r^2}\mathcal D^jF_{ji}=0,
 \label{eq:YMradialevolution}
\end{equation}
where $\mathcal D$ includes the Levi-Civita connection of the unit sphere and
the gauge connection, and the index $j$ is raised with the round metric of
the unit sphere. Repeated radial differentiation shows inductively that all
derivatives of $A_i$ vanish at $r=1$. Extending the connection by zero to an
exterior collar therefore gives a smooth Yang-Mills connection. Every smooth
connection admits a local Coulomb gauge~\cite{Uhlenbeck,DK}; in it, the
Yang-Mills equations for the flat metric form an elliptic system with
analytic coefficients, whose solutions are real analytic~\cite{Morrey}.
Hence the gauge-invariant function $\sum_{a,\mu,\nu}(F^a_{\mu\nu})^2$ is real
analytic on the connected extended ball. It vanishes in the exterior collar,
so it vanishes throughout the ball. For a real connection with a positive
invariant inner product, this implies $F=0$.
\end{proof}

For self-dual connections it is enough that the tangential components
$F_{jk}$ vanish on $S^3$: by \eqref{eq:localdual}, self-duality then gives
$F_{1i}=\tfrac12\lc_{1ijk}F_{jk}=0$ as well. 

\begin{theorem}[No smooth real non-flat saddles]\label{thm:main}
Let $A$ be a real self-dual connection, smooth on the closed conformal ball of Euclidean AdS$_4$, for a compact gauge group with a positive invariant inner product, and let $g_+^{-2}\ne0$. If there is a smooth, possibly complex, anti-self-dual adjoint tensor $B_{\mu\nu}$ such that $(A,B)$ solves \eqref{eq:critical}, then $A$ is flat.
\end{theorem}

\begin{proof}
Apply Theorem~\ref{thm:stationary} to the family \eqref{eq:dilatation} at $\lambda=1$. The family is defined only for $\lambda\le1$, but a one-sided derivative suffices. Its tangent $a_\mu=A_\mu+y^\nu\partial_\nu A_\mu$ is smooth on $\bar\Ball$, so no extension of $A$ beyond the ball is needed. By \eqref{eq:dQdlambda},
\begin{equation}
 0=\left.\frac{\dd Q[A_\lambda]}{\dd\lambda}\right|_{1^-}
 =\frac1{32\pi^2}\int_{S^3}\sum_{a,\mu,\nu}(F^a_{\mu\nu})^2\,\dd\Omega_3 .
 \label{eq:positiveboundary}
\end{equation}
The integrand is continuous and nonnegative, so it vanishes at every point of $S^3$, and so does every component of $F$ there. Proposition~\ref{prop:uc} gives $F=0$ everywhere.
\end{proof}

Reality of $A$ enters only through the positivity used in \eqref{eq:positiveboundary}.

\begin{corollary}\label{cor:B}
Under the hypotheses of Theorem~\ref{thm:main}, $(A,B)$ is gauge equivalent to $(0,0)$.
\end{corollary}

\begin{proof}
A flat connection on the simply connected ball is pure gauge, $A_\mu=U^{-1}\partial_\mu U$ with $U$ smooth on $\bar\Ball$. After the gauge transformation by $U$, which maps $B_{\mu\nu}\to UB_{\mu\nu}U^{-1}$ and preserves all conditions in \eqref{eq:critical}, we have $A_\mu=0$. In Cartesian coordinates the equations become
\begin{equation*}
 \partial_\mu B_{\mu\nu}=0,\qquad \star B=-B,\qquad n_\mu B_{\mu\nu}\big|_{S^3}=0 ,
\end{equation*}
the last because $F=0$. Since $\star B=-B$, the first equation also says $\partial_\mu(\star B)_{\mu\nu}=0$, that is, $\lc_{\nu\mu\rho\sigma}\partial_\mu B_{\rho\sigma}=0$, or equivalently $\partial_\mu B_{\nu\rho}+\partial_\nu B_{\rho\mu}+\partial_\rho B_{\mu\nu}=0$. On the star-shaped ball this implies (Poincar\'e's lemma) that $B$ is a curl,
\begin{equation*}
 B_{\mu\nu}=\partial_\mu\zeta_\nu-\partial_\nu\zeta_\mu,\qquad
 \zeta_\nu(y)=\int_0^1t\,y^\mu B_{\mu\nu}(ty)\,\dd t ,
\end{equation*}
with $\zeta_\nu$ smooth on $\bar\Ball$. Then, even for complex $B$,
\begin{equation}
 \sum_{a,\mu,\nu}\int_\Ball\dd^4y\,\big|B^a_{\mu\nu}\big|^2
 =2\int_\Ball\dd^4y\,\overline{B^a_{\mu\nu}}\,\partial_\mu\zeta^a_\nu
 =2\oint_{S^3}\dd\Sigma\,n_\mu\overline{B^a_{\mu\nu}}\,\zeta^a_\nu
 -2\int_\Ball\dd^4y\,\big(\partial_\mu\overline{B^a_{\mu\nu}}\big)\zeta^a_\nu=0 .
 \label{eq:Bstokes}
\end{equation}
The boundary term vanishes by the boundary condition and the bulk term by the divergence condition. Hence $B=0$.
\end{proof}

The boundary condition is needed here as well. A nonzero constant anti-self-dual tensor, such as $B_{12}=-B_{34}=\text{const}$, satisfies the bulk equations on the ball.

One can also see the dilation obstruction directly from the action. Keep the coordinates, metric, domain and $B$ fixed, and vary the gauge field by
\begin{equation}
 \delta A_\mu=-(A_\mu+y^\rho\partial_\rho A_\mu),
 \qquad \delta B=0 .
 \label{eq:activedilation}
\end{equation}
A direct calculation gives $\delta F_{\mu\nu}=-(2+y^\rho\partial_\rho)F_{\mu\nu}$. Since the flat Hodge star has constant coefficients, $\delta F^-=0$ at a self-dual field. The BF term therefore has zero first variation for any $B$. With $F^2=\sum_{a,\mu,\nu}(F^a_{\mu\nu})^2$, the remaining term gives
\begin{align}
 \delta S_0
 &=\frac1{2g_+^2}\int_\Ball F^a_{\mu\nu}\delta F^a_{\mu\nu}\,\dd^4y
 =-\frac1{4g_+^2}\int_\Ball\partial_\rho(y^\rho F^2)\,\dd^4y
 \nonumber\\
 &=-\frac1{4g_+^2}\int_{S^3}F^2\,\dd\Omega_3 .
 \label{eq:activevariation}
\end{align}
Stationarity again gives \eqref{eq:positiveboundary}. The variation \eqref{eq:activedilation} is the tangent at $t=0$ to $A_\mu(t,y)=e^{-t}A_\mu(e^{-t}y)$, which is \eqref{eq:dilatation} with $\lambda=e^{-t}$. For a real non-flat self-dual field, Proposition~\ref{prop:uc} makes the boundary integral strictly positive. The action decreases in this direction if $g_+^{-2}$ is real and positive. For any nonzero complex $g_+^{-2}$ its first variation is still nonzero, which is enough to exclude a saddle.

\section{Exact solutions and the 2009 charge correction}\label{sec:example}

We now apply the result to the $SO(3)$-symmetric solutions of Ref.~\cite{ST2009}. Their vortex description also allows us to compare the flux used in that paper with the four-dimensional Pontryagin charge.

\subsection{Reduction to vortices}

\begin{figure}[t]
\centering
\begin{tikzpicture}[scale=2.6]
\fill[gray!8] (0,-1) arc (-90:90:1) -- cycle;
\draw[thick] (0,-1) arc (-90:90:1);
\draw[thick,dashed] (0,-1.15) -- (0,1.15);
\draw[->] (0,0) -- (1.35,0) node[right] {$q$};
\draw[->] (0,0) -- (0,1.3) node[above] {$w$};
\draw[densely dotted,domain=-53.13:53.13,samples=60]
  plot ({-0.75+1.25*cos(\x)},{1.25*sin(\x)});
\fill (0.5,0) circle (0.022);
\node[below right] at (0.5,0) {vortex};
\node at (0.83,0.86) [right] {$|u|=1$: boundary of AdS$_4$};
\node at (-0.06,0.62) [left] {axis $\rho=0$};
\fill (0,1) circle (0.02) node[left] {$t=+\infty$};
\fill (0,-1) circle (0.02) node[left] {$t=-\infty$};
\node at (0.68,-0.42) {$D_+$};
\end{tikzpicture}
\caption{The orbit space $D_+$ of $SO(3)$ acting on Euclidean AdS$_4$. With the metric $\gamma$ it is a hyperbolic half-plane bounded by the geodesic $|u|=1$, the conformal boundary of AdS$_4$. The dotted arc is a curve at fixed distance $d$ from the boundary, defined in \eqref{eq:d}; AdS time translations move the vortex along it.}
\label{fig:orbit}
\end{figure}
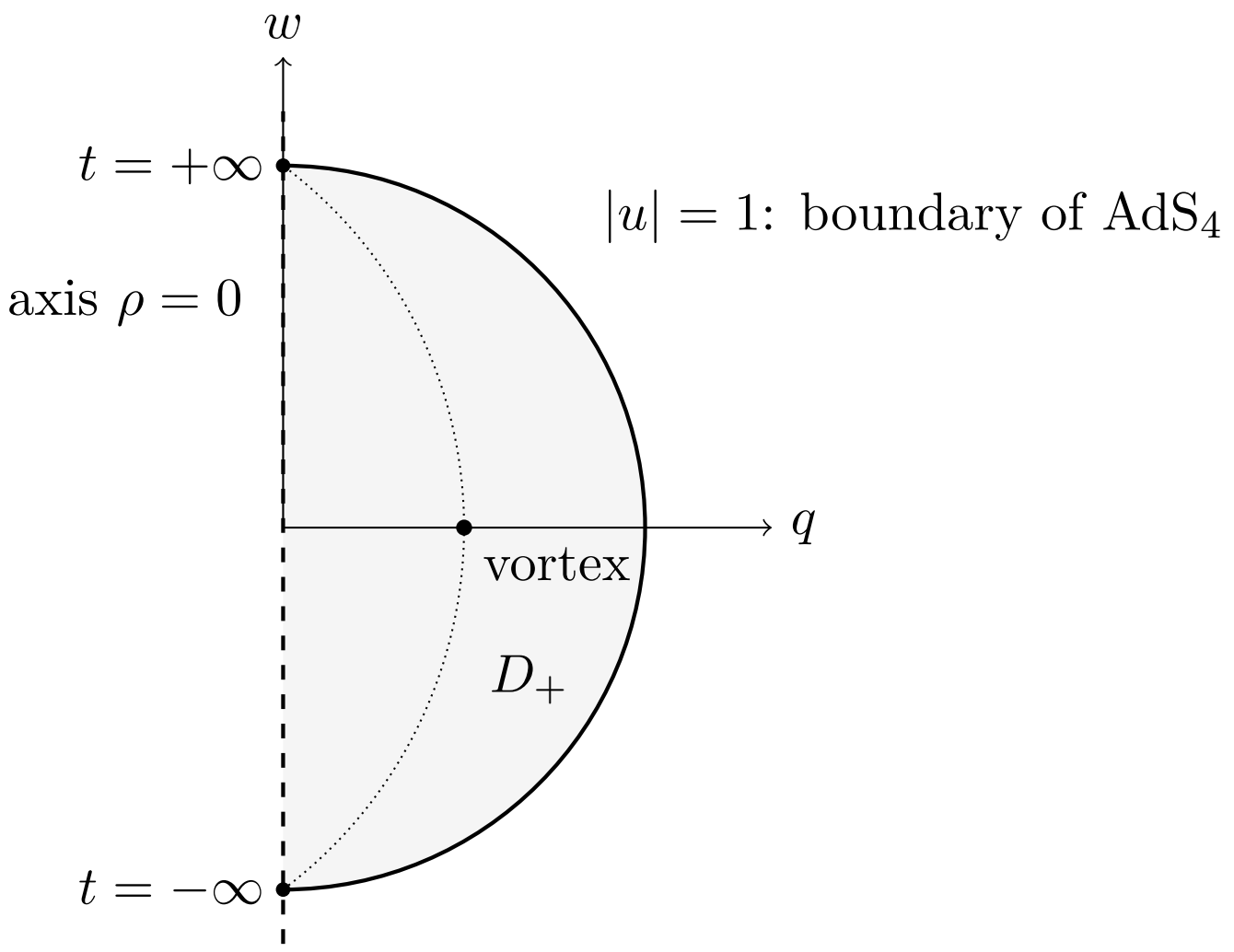

Let $SO(3)$ rotate $(y^1,y^2,y^3)$ and write $y=(q\,\hat n,w)$ with $q\ge0$, $\hat n\in S^2$ and $w=y^4$. The flat metric is $q^2(\gamma+\dd\Omega_2^2)$, where $\dd\Omega_2^2$ is the round metric on the unit two-sphere and
\begin{equation*}
 \gamma=\frac{\dd q^2+\dd w^2}{q^2}.
\end{equation*}
The space of orbits, the half-plane $q>0$ with complex coordinate $u=q+iw$, therefore carries the hyperbolic metric $\gamma$. Since self-duality is conformally invariant, $SO(3)$-symmetric self-dual fields reduce to vortices on the hyperbolic plane~\cite{Witten1977,Tekin2000,MantonRink}; related constructions exist on other curved backgrounds~\cite{NianQian}. The vortex equations, in turn, reduce to the Liouville equation~\cite{Witten1977}, one of many reductions of the self-duality equations to lower-dimensional integrable systems~\cite{Ward1985}. On AdS$_4$ only the domain changes. The conformal boundary $|y|=1$ becomes the semicircle $|u|=1$, a geodesic of $\gamma$, and the orbit space is the half-disk
\begin{equation*}
 D_+=\{u=q+iw:\ q>0,\ |u|<1\},
\end{equation*}
a hyperbolic half-plane bounded by that geodesic (Fig.~\ref{fig:orbit}). The edge $q=0$ is the axis on which the two-sphere orbits shrink to points. It lies inside AdS$_4$ and is not a boundary: there, one imposes regularity, not a boundary condition.

In global coordinates $\dd s^2=(1+\rho^2/\ell^2)\dd t^2+\dd\rho^2/(1+\rho^2/\ell^2)+\rho^2\dd\Omega_2^2$, with $t\in\mathbb R$ and $\rho\ge0$,
\begin{equation}
 u=\tan\frac{\arctan(\rho/\ell)+it/\ell}{2}.
 \label{eq:global}
\end{equation}
The center $\rho=0$ maps to the axis, the boundary $\rho=\infty$ to $|u|=1$, and $t\to\pm\infty$ to $u\to\pm i$. Time translations act on $u$ as M\"obius maps, fixing $\pm i$. They are hyperbolic isometries that move points along arcs at a constant distance from the boundary geodesic.

Let $\vartheta,\varphi$ be the polar angles of $\hat n$, and put $T_r=\hat n^aT_a$, $T_\vartheta=\partial_\vartheta\hat n^aT_a$ and $T_\varphi=(\sin\vartheta)^{-1}\partial_\varphi\hat n^aT_a$. In the coordinates $(q,w,\vartheta,\varphi)$, Witten's ansatz~\cite{Witten1977} gives the components
\begin{equation}
 A_q=\mathfrak a_q\,T_r,\qquad A_w=\mathfrak a_w\,T_r,\qquad
 A_\vartheta=\phi_1T_\vartheta+(\phi_2-1)T_\varphi,\qquad
 A_\varphi=\sin\vartheta\left[\phi_1T_\varphi-(\phi_2-1)T_\vartheta\right],
 \label{eq:ansatz}
\end{equation}
where $(\mathfrak a_q,\mathfrak a_w)$ is an Abelian gauge field and $\phi=\phi_1+i\phi_2$ a complex scalar on the orbit space, all functions of $(q,w)$ only. The vacuum is $\phi=i$, $\mathfrak a_q=\mathfrak a_w=0$. Let $D_q\phi=\partial_q\phi+i\mathfrak a_q\phi$, $D_w\phi=\partial_w\phi+i\mathfrak a_w\phi$, $\partial_{\bar u}=\frac12(\partial_q+i\partial_w)$ and $\mathfrak a_{\bar u}=\frac12(\mathfrak a_q+i\mathfrak a_w)$. The field \eqref{eq:ansatz} is self-dual if and only if
\begin{equation}
 D_{\bar u}\phi\equiv\partial_{\bar u}\phi+i\mathfrak a_{\bar u}\phi=0,\qquad
 \partial_q\mathfrak a_w-\partial_w\mathfrak a_q=-\frac{1-|\phi|^2}{q^2}.
 \label{eq:vortex}
\end{equation}
The gauge transformations $e^{\chi T_r}$, with $\chi$ a real function on the orbit space, preserve the ansatz and act as $\phi\to e^{-i\chi}\phi$, $\mathfrak a_q\to\mathfrak a_q+\partial_q\chi$, $\mathfrak a_w\to\mathfrak a_w+\partial_w\chi$. For a smooth field on the axis, $1-|\phi|^2=O(q^2)$ there.

Two properties of \eqref{eq:vortex} are used below. First, on a simply connected region, solutions with the same $|\phi|$ are gauge equivalent. Locally $\phi=e^{-h}\Psi$ with $\partial_{\bar u}h=i\mathfrak a_{\bar u}$ and $\Psi$ holomorphic, so the zeros of $\phi$ are isolated and their multiplicities are fixed by $|\phi|$. If two solutions $\phi$ and $\phi'$ have $|\phi'|=|\phi|$, the ratio $\phi'/\phi$ is a smooth function of unit modulus, hence equal to $e^{-i\chi}$ with $\chi$ real and smooth, and the first equation gives $\mathfrak a'_q=\mathfrak a_q+\partial_q\chi$ and $\mathfrak a'_w=\mathfrak a_w+\partial_w\chi$. Second, for every holomorphic map $\Phi$ from the half-plane $q>0$ into the unit disk, there is a solution~\cite{Witten1977,Tekin2000} with
\begin{equation}
 |\phi|=\frac{2q\,|\Phi'(u)|}{1-|\Phi(u)|^2},\qquad \Phi'=\frac{\dd\Phi}{\dd u},
 \label{eq:witten}
\end{equation}
whose zeros, the vortices, are the critical points of $\Phi$. A M\"obius map $\Phi$ is an isometry onto the disk and gives $|\phi|=1$, the vacuum. A product of $k$ M\"obius factors, a finite Blaschke product, gives $k-1$ vortices.

\subsection{The one-vortex family is a restricted BPST instanton}

The first family studied in Ref.~\cite{ST2009} is
\begin{equation}
 \Phi(u)=\left(\frac{u_0-u}{\bar u_0+u}\right)^2,\qquad u_0=\alpha+i\beta,\quad\alpha>0,
 \label{eq:Phia}
\end{equation}
with a single vortex at $u=u_0$. The solution with $Q=1/2$ in that paper is the case $u_0=1$. On the full half-plane this is the charge-one solution of Witten's construction, the BPST instanton~\cite{BPST}: the real part of the vortex position is the instanton scale, and the imaginary part is its position along the axis~\cite{Witten1977,Tekin2000}. The 2009 family is therefore the BPST instanton of scale $\alpha$ centered at $b=(0,0,0,\beta)$, restricted to the ball. To see this directly, write the BPST field in regular gauge, in Cartesian components,
\begin{equation}
 A^{\rm BPST}_\mu=\frac{2\eta^a_{\mu\nu}(y-b)^\nu}{|y-b|^2+\alpha^2}\,T_a ,
 \label{eq:BPST}
\end{equation}
with 't~Hooft's symbols~\cite{tHooft} in the convention $\eta^a_{ij}=\epsilon_{aij}$, $\eta^a_{i4}=-\eta^a_{4i}=\delta_{ai}$, $\eta^a_{44}=0$. Its curvature is $F^a_{\mu\nu}=-4\alpha^2\eta^a_{\mu\nu}/(|y-b|^2+\alpha^2)^2$, which is self-dual in the orientation of Sec.~\ref{sec:ball}; by \eqref{eq:flatdensity} its charge density is $\dens=6\alpha^4/[\pi^2(|y-b|^2+\alpha^2)^4]$. The BPST connection is smooth on all of $\mathbb R^4$ and, for $b$ on the axis, $SO(3)$-symmetric. Inserting $y=(q\hat n,w)$ and transforming the components to $(q,w,\vartheta,\varphi)$ brings it to the form \eqref{eq:ansatz} with
\begin{equation}
 \phi_{\rm BPST}=\frac{i(\alpha^2-v^2)}{\alpha^2+|v|^2},\qquad
 \mathfrak a_q=\frac{2(w-\beta)}{\alpha^2+|v|^2},\qquad
 \mathfrak a_w=-\frac{2q}{\alpha^2+|v|^2},
 \qquad v=u-i\beta .
 \label{eq:BPSTred}
\end{equation}
These satisfy \eqref{eq:vortex}, equal $\phi=i$ on the axis, vanish at $u=u_0$, and have
\begin{equation}
 |\phi_{\rm BPST}|^2=1-\frac{4\alpha^2q^2}{(\alpha^2+|v|^2)^2}
 =\frac{4q^2|\Phi'|^2}{(1-|\Phi|^2)^2}.
 \label{eq:samemod}
\end{equation}
The two reduced solutions have the same $|\phi|$ and $D_+$ is simply connected, so they are related by a residual gauge transformation on all of $D_+$. Hence:

\begin{proposition}\label{prop:ident}
For every $u_0$ with $\mathrm{Re}\,u_0>0$, the self-dual field of Ref.~\cite{ST2009} built from \eqref{eq:Phia} is gauge equivalent, off the axis, to the BPST instanton \eqref{eq:BPST} restricted to the ball. In particular, the $Q=1/2$ solution is the unit-scale BPST instanton centered at the origin of the ball.
\end{proposition}

The words ``off the axis'' matter. The gauge transformation is $e^{\chi T_r}$, and $T_r=\hat n^aT_a$ has no limit at $q=0$, so a smooth $\chi$ on $D_+$ does not by itself give a smooth four-dimensional gauge transformation. In the holomorphic gauge of Ref.~\cite{ST2009} the phase of $\phi$ follows that of $\Phi'$, which varies along the axis, whereas regularity there requires $\phi=i$. That gauge is therefore singular along a line inside AdS$_4$. The BPST formula, by contrast, is smooth on the closed ball and is a regular representative of the same orbit-space solution. We compute the charge from this regular BPST representative.

\subsection{The charge}

Reference~\cite{ST2009} computed the charge as the flux of the Abelian field,
\begin{equation}
 \Vf=\frac1{2\pi}\int_{D_+}(1-|\phi|^2)\,\frac{\dd q\,\dd w}{q^2}
 =-\frac1{2\pi}\int_{D_+}\big(\partial_q\mathfrak a_w-\partial_w\mathfrak a_q\big)\,\dd q\,\dd w .
 \label{eq:flux}
\end{equation}
On the full hyperbolic plane, for Witten's solutions built from finite Blaschke products, $\Vf$ counts the zeros of $\Phi'$ with multiplicity and equals the instanton number~\cite{Witten1977,Tekin2000}. On the half-disk it does not. Write $\mathrm{vol}_\gamma=\dd q\,\dd w/q^2$, $|D\phi|^2_\gamma=q^2(|D_q\phi|^2+|D_w\phi|^2)$ and $\Delta_\gamma=q^2(\partial_q^2+\partial_w^2)$ for the area element, the gradient norm and the Laplacian of $\gamma$. A direct decomposition of the four-dimensional curvature (Appendix~\ref{app:4d}) gives, for a self-dual field,
\begin{equation}
 8\pi^2Q=4\pi\int_{D_+}\left[\,|D\phi|^2_\gamma+(1-|\phi|^2)^2\right]
 \mathrm{vol}_\gamma ,
 \label{eq:redaction}
\end{equation}
and the vortex equations give the pointwise identity (Appendix~\ref{app:id})
\begin{equation}
 |D\phi|_\gamma^2+(1-|\phi|^2)^2=(1-|\phi|^2)+\tfrac12\Delta_\gamma|\phi|^2 .
 \label{eq:pointwise}
\end{equation}
In two dimensions the Laplacian is conformally covariant, $\Delta_\gamma f\,\mathrm{vol}_\gamma=(\partial_q^2+\partial_w^2)f\,\dd q\,\dd w$, so the last term integrates to a flat-space boundary integral:
\begin{equation}
 Q=\Vf+\Bd,\qquad
 \Bd=\frac1{4\pi}\oint_{|u|=1}\partial_\perp|\phi|^2\,\dd s ,
 \label{eq:Qcorrected}
\end{equation}
where $\partial_\perp$ and $\dd s$ are the flat outward normal derivative and arc length on the semicircle. The axis does not contribute, because $1-|\phi|^2=O(q^2)$ there. The 2009 reduction set the boundary term to zero~\cite{ST2009}. On the AdS orbit space, however, the semicircle $|u|=1$ is the conformal boundary, and the normal derivative of $|\phi|^2$ need not vanish there. This is the source of the difference between the vortex flux and the four-dimensional charge.

For the family \eqref{eq:Phia} all three quantities can be computed exactly. AdS time translations are isometries that commute with $SO(3)$. On the orbit space, they act as isometries of $\gamma$ and move $u_0$ along curves at constant signed hyperbolic distance $d$ from the boundary geodesic (Fig.~\ref{fig:orbit}); they map the solution with a vortex at $u_0$ to the solution with the displaced vortex, up to gauge, since $|\phi|$ in \eqref{eq:witten} is unchanged when $\Phi$ is composed with a M\"obius map and multiplied by a phase. A time translation maps the closed ball to itself, preserving orientation, so it leaves $Q$ unchanged by the metric-independent form of \eqref{eq:Q}; it leaves $\Vf$ unchanged because $\Vf$ is the integral of $1-|\phi|^2$ against the area element of $\gamma$; hence, it also leaves $\Bd=Q-\Vf$ unchanged. The charges, therefore, depend on $u_0$ only through
\begin{equation}
 X\equiv\sinh d=\frac{1-|u_0|^2}{2\,\mathrm{Re}\,u_0}=\frac{\ell}{\rho_0}.
 \label{eq:d}
\end{equation}
The last expression, with $\rho_0$ the global radial coordinate of the vortex, applies when the vortex lies inside AdS$_4$ ($d>0$). For $|u_0|>1$ the vortex lies outside the physical orbit space and $d<0$, but $d$ is still a good label for the restricted BPST field. In the centered case $\beta=0$ one has $1-|\phi|^2=4\alpha^2q^2/(\alpha^2+|u|^2)^2$, and elementary integrals give
\begin{equation}
 \Vf=\frac1{1+\alpha^2},\qquad
 \Bd=\frac{\alpha^2(1-\alpha^2)}{(1+\alpha^2)^3},\qquad
 Q=\frac{1+3\alpha^2}{(1+\alpha^2)^3},
 \label{eq:Qcentred}
\end{equation}
which is \eqref{eq:introcharges}. The value of $Q$ agrees with the direct integration of the BPST charge density, $Q=12\alpha^4\int_0^1r^3\,\dd r/(r^2+\alpha^2)^4$ for $b=0$. The boundary term can also be computed from its definition: on $|u|=1$, $\partial_\perp|\phi|^2=8\alpha^2(1-\alpha^2)q^2/(1+\alpha^2)^3$, whose integral over the semicircle gives the same $\Bd$. With $\tau=\tanh d=(1-\alpha^2)/(1+\alpha^2)$ the results become
\begin{equation}
 \Vf=\frac{1+\tau}{2},\qquad
 \Bd=\frac{\tau(1-\tau^2)}{4},\qquad
 Q=\frac{(1+\tau)^2(2-\tau)}{4},\qquad \tau=\tanh d .
 \label{eq:QofT}
\end{equation}
These formulas hold for the whole family. The charge formula holds, in fact, for a BPST instanton of any center $b\in\mathbb R^4$ and scale $\alpha$, with
\begin{equation}
 X=\sinh d=\frac{1-|b|^2-\alpha^2}{2\alpha}.
 \label{eq:X4}
\end{equation}
A rotation of $\mathbb R^4$ about the origin is also an orientation-preserving isometry of AdS$_4$, so it leaves $Q$ unchanged; it moves $b$ onto the axis without changing $|b|$. For $b=(0,0,0,\beta)$, Proposition~\ref{prop:ident} identifies the field with the 2009 field with $u_0=\alpha+i\beta$, and \eqref{eq:d} gives \eqref{eq:X4}.

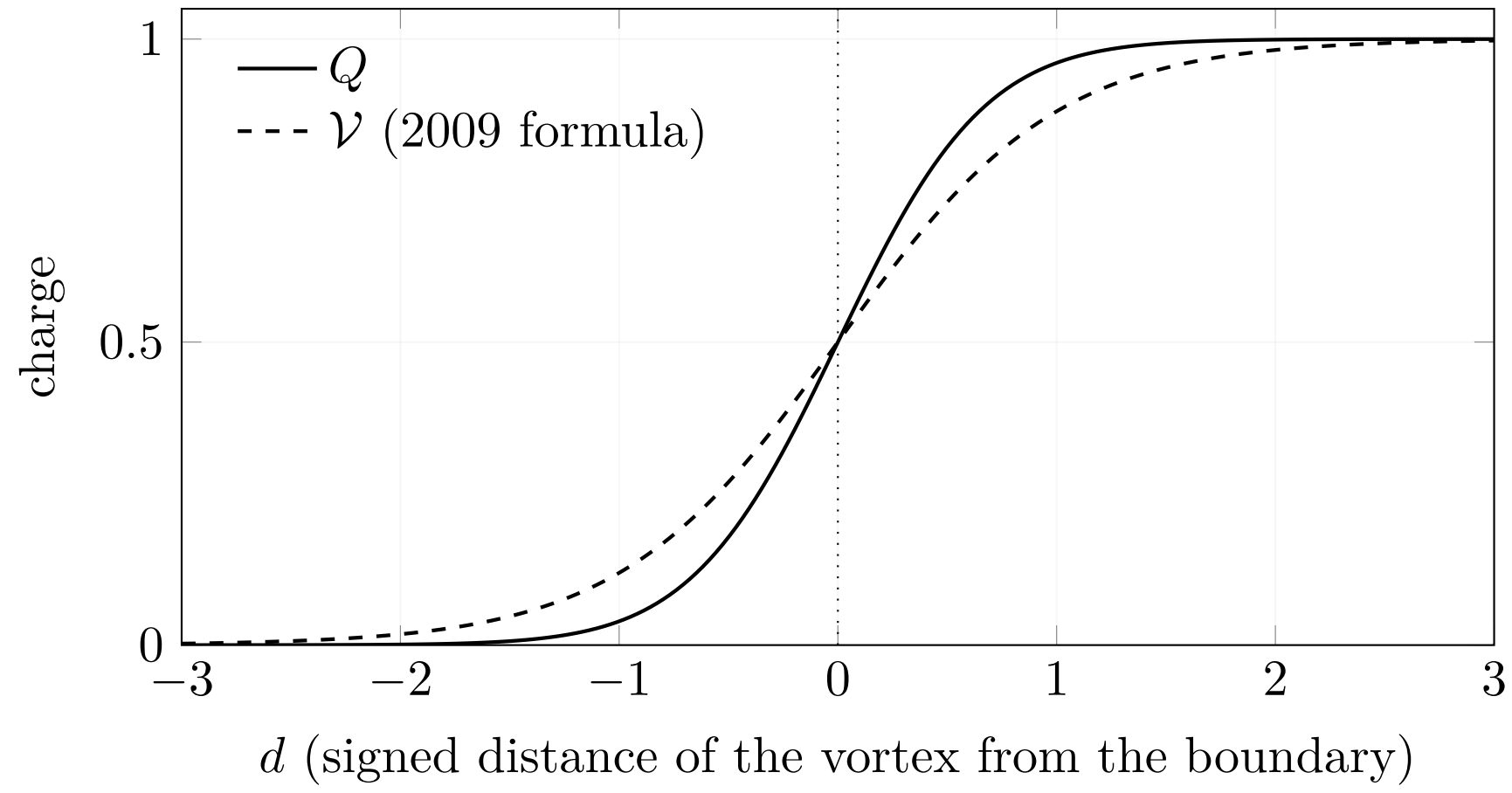
\begin{figure}[t]
\centering
\begin{tikzpicture}
\begin{axis}[
  width=0.72\textwidth,height=0.4\textwidth,
  xlabel={$d$ (signed distance of the vortex from the boundary)},
  ylabel={charge},
  xmin=-3,xmax=3,ymin=0,ymax=1.05,
  domain=-3:3,samples=200,
  legend style={at={(0.03,0.97)},anchor=north west,draw=none,fill=none},
  legend cell align=left,
  grid=both,grid style={opacity=0.12}]
\addplot[thick] {(1+tanh(x))^2*(2-tanh(x))/4};
\addlegendentry{$Q$}
\addplot[thick,dashed] {(1+tanh(x))/2};
\addlegendentry{$\Vf$ (2009 formula)}
\addplot[thin,dotted] coordinates {(0,0) (0,1.05)};
\end{axis}
\end{tikzpicture}
\caption{The Pontryagin charge $Q$ and the vortex flux $\Vf$ of the one-vortex family against the signed distance $d$ between the vortex and the boundary of AdS$_4$. They coincide only at $d=0$ and as $d\to\pm\infty$.}
\label{fig:charges}
\end{figure}

Figure~\ref{fig:charges} shows the result. A vortex deep inside AdS$_4$ ($d\to+\infty$) belongs to a small instanton, and $Q\to1$. A vortex far outside the physical region ($d\to-\infty$) belongs to an instanton whose core lies outside the ball or whose scale exceeds it, and $Q\to0$. For $d<0$ the field has no zero of $\phi$ inside AdS$_4$ at all, yet $Q\ne0$. The boundary term $\Bd$ vanishes exactly at $d=0$, where the vortex lies on the boundary geodesic. The configuration is then symmetric under reflection in that geodesic, so $\partial_\perp|\phi|^2=0$. This is the case $|u_0|=1$, which contains the 2009 solution $u_0=1$ and its time translates, and for these, the 2009 value $Q=1/2$ is correct. For all other one-vortex parameters, the integral evaluated in Ref.~\cite{ST2009} is $\Vf$, not the four-dimensional charge $Q$. The difference reaches $|\Bd|=1/(6\sqrt3)\approx0.096$ at $\tau=\pm1/\sqrt3$. Even at $d=0$ the slopes differ: $\dd Q/\dd d=\frac34\,\mathrm{sech}^4d$, whereas $\dd\Vf/\dd d=\frac12\,\mathrm{sech}^2d$.

Equation~\eqref{eq:QofT}, together with \eqref{eq:d}, replaces the numerical charge calculation in Fig.~1 of Ref.~\cite{ST2009}. It gives the full dependence on $u_0=\alpha+i\beta$ through $d$. For example, at $\alpha=1$, increasing $\beta$ from $0$ to $1$ lowers $Q$ from $1/2$ to about $0.19$ and $\Vf$ from $1/2$ to about $0.28$.

The value $Q=1/2$ at $d=0$ has a four-dimensional explanation. By \eqref{eq:X4}, $d=0$ means $|b|^2+\alpha^2=1$. For such an instanton, the measure $\dens\,\dd^4y$ is invariant under the inversion $y\to y/|y|^2$ in the boundary sphere, which exchanges the inside and the outside of the ball. Half of the total charge, therefore, lies inside.

The centered value of $Q$ can also be read off from the boundary connection alone, through \eqref{eq:CS}. For $b=0$ the BPST field \eqref{eq:BPST} is
\begin{equation}
 A_\mu=h(r)\,\Theta_\mu,\qquad h(r)=\frac{r^2}{r^2+\alpha^2},\qquad
 \Theta_\mu=\frac{2\eta^a_{\mu\nu}y^\nu}{r^2}\,T_a,\qquad r=|y|.
 \label{eq:centered}
\end{equation}
The field $\Theta_\mu$ is defined only for $r\ne0$, but $h\Theta_\mu$ is smooth at the origin. It has no radial component, $y^\mu\Theta_\mu=0$, and it is pure gauge,
\begin{equation}
 \partial_\mu\Theta_\nu-\partial_\nu\Theta_\mu+[\Theta_\mu,\Theta_\nu]=0 .
 \label{eq:Thetaflat}
\end{equation}
Using \eqref{eq:Thetaflat} to eliminate the commutator, the field strength of $A_\mu=h\Theta_\mu$ is
\begin{equation*}
 F_{\mu\nu}=\frac{h'}{r}\big(y_\mu\Theta_\nu-y_\nu\Theta_\mu\big)
 +\big(h-h^2\big)\big(\partial_\mu\Theta_\nu-\partial_\nu\Theta_\mu\big).
\end{equation*}
On $S^3$ the first term has one index along the normal, so the tangential field strength of the boundary connection $h_0\Theta_\mu$, with $h_0=h(1)=1/(1+\alpha^2)$, is $(h_0-h_0^2)(\partial_\mu\Theta_\nu-\partial_\nu\Theta_\mu)$. It vanishes only for $h_0=0$ or $1$. The normal component $n_\mu K^\mu$ of the current \eqref{eq:Kcurrent} contains only tangential components of $A$ and tangential derivatives, so it depends on $h$ only through $h_0$. A short calculation with $A_\mu=h_0\Theta_\mu$ gives $n_\mu K^\mu=8(3h_0^2-2h_0^3)$ at every point of $S^3$. Since the unit sphere has area $2\pi^2$, Eq.~\eqref{eq:CS} gives
\begin{equation}
 Q=3h_0^2-2h_0^3=\frac{1+3\alpha^2}{(1+\alpha^2)^3},
 \label{eq:Qh0}
\end{equation}
in agreement with \eqref{eq:Qcentred}. The charge is a polynomial in the boundary datum $h_0$ and takes every value between the flat points $h_0=0$ and $h_0=1$; the unit-scale instanton has $h_0=\tfrac12$ and $Q=\tfrac12$.

The small-size limit needs care. If the instanton center stays strictly inside the ball and $\alpha\to0$, the curvature concentrates at the center and $Q\to1$; the field does not converge to any smooth configuration. With the center fixed outside the ball, $Q\to0$ and the field inside tends to the vacuum. A center approaching the sphere at a rate comparable to $\alpha$ gives intermediate values. The zero-size limit is therefore not the smooth connection obtained by setting $\alpha=0$ in the gauge potential. In particular, the value $Q=0$ that Ref.~\cite{ST2009} assigns to $\alpha=0$ refers to the configuration with $\alpha$ set to zero from the start, not to the limit: for $|\beta|<1$ the limit $\alpha\to0$ gives $Q\to1$, and also $\Vf\to1$.

The family also illustrates Theorem~\ref{thm:main}. The dilation \eqref{eq:dilatation} acts on the family as $u_0\to u_0/\lambda$, that is, $(\alpha,\beta)\to(\alpha,\beta)/\lambda$. On the orbit space, it is the isometry $u\to u/\lambda$ that moves the vortex relative to the boundary geodesic. For centered instantons, it acts as $d\to d+\log\lambda$, and in general, $X=\sinh d$ changes at the rate
\begin{equation}
 \left.\frac{\partial X}{\partial\lambda}\right|_{\lambda=1}=\frac{1+\alpha^2+\beta^2}{2\alpha}>0 .
 \label{eq:Xrate}
\end{equation}
Since $\dd Q/\dd d>0$ everywhere, the charge of every member of the family changes under dilation, and no member admits a $B$ with the properties required by Theorem~\ref{thm:stationary}. For a centered instanton, this can be checked without the distance variable:
\begin{equation}
 \frac{\dd Q}{\dd\alpha}=-\frac{12\alpha^3}{(1+\alpha^2)^4}\ne0,
 \qquad
 \left.\frac{\dd Q[A_\lambda]}{\dd\lambda}\right|_{1^-}
 =\frac{12\alpha^4}{(1+\alpha^2)^4}>0 .
 \label{eq:explicitvariation}
\end{equation}
The second expression also follows by evaluating the boundary integral in \eqref{eq:positiveboundary}. In boundary language, along the boundary connections $h_0\Theta_\mu$ the on-shell action \eqref{eq:onshell} equals $(8\pi^2/g_+^2)(3h_0^2-2h_0^3)$ by \eqref{eq:Qh0}, and its derivative $(8\pi^2/g_+^2)\,6h_0(1-h_0)$ vanishes only at the flat ends $h_0=0$ and $h_0=1$.

The failure of stationarity does not depend on the charge correction. Even the uncorrected flux varies, $\dd\Vf/\dd\alpha=-2\alpha/(1+\alpha^2)^2$, so it has no stationary size either. At the symmetric half-charge point, with primes denoting $\dd/\dd\alpha$,
\begin{equation}
 \Bd(1)=0,\qquad \Bd'(1)=-\tfrac14,\qquad
 Q'(1)=-\tfrac34,\qquad\Vf'(1)=-\tfrac12 .
 \label{eq:surfacevariation}
\end{equation}
A surface term that vanishes for one field need not have a vanishing variation there. The half-charge value survives, but it does not give a saddle of the HRSS boundary problem.

\subsection{The higher-power family}

The second family of Ref.~\cite{ST2009} is $\Phi=\Phi_v^n$, $n\ge2$, with $\Phi_v(u)=(1-u)/(1+u)$. On the full half-plane, it has a total instanton charge $n-1$~\cite{Witten1977}; its single vortex, of multiplicity $n-1$, sits at $u=1$ on the boundary geodesic. For $n=2$ it is the one-vortex solution with $u_0=1$. Reflection in the geodesic $u\to1/\bar u$ maps $\Phi_v^n$ to $(-1)^n\overline{\Phi_v^n}$ and leaves $|\phi|$ invariant. Hence, $\Bd=0$. The reflection is an isometry of $\gamma$ that exchanges $D_+$ with the rest of the half-plane, so exactly half of the full-plane flux $n-1$ lies in $D_+$. The axis does not contribute to \eqref{eq:Qcorrected} either. The map $\Phi_v^n$ is a finite Blaschke product, a power of the Cayley map $\Phi_v$, so $|\Phi_v^n|=1$ and $(\Phi_v^n)'\neq0$ on the axis. By the Schwarz reflection $\Phi(-\bar u)=1/\overline{\Phi(u)}$, $|\phi|$ is even in $q$ and equals $1$ at $q=0$, so $1-|\phi|^2=O(q^2)$. Hence, for every $n\ge2$,
\begin{equation}
 Q=\Vf=\frac{n-1}{2}.
 \label{eq:Qpower}
\end{equation}
The four-dimensional field is smooth on the closed ball, although the holomorphic gauge of Ref.~\cite{ST2009} is not. Away from the vortex, the gauge $\phi=i|\phi|$, $\mathfrak a_q=-\partial_w\log|\phi|$, $\mathfrak a_w=\partial_q\log|\phi|$ solves \eqref{eq:vortex}; since $|\phi|$ is real analytic and even in $q$ with $|\phi|=1$ on the axis, $\mathfrak a_q=O(q^2)$, $\mathfrak a_w=O(q)$, and $|\phi|-1=O(q^2)$, which is what the smoothness of the Cartesian components of \eqref{eq:ansatz} on the axis requires. Near the vortex, with $z=\Phi_v(u)$ and $\Lambda=|\phi|/|z|^{n-1}=n/\sum_{j=0}^{n-1}|z|^{2j}$, which is smooth and positive, the gauge $\phi=i\Lambda z^{n-1}$, $\mathfrak a_q=-\partial_w\log\Lambda$, $\mathfrak a_w=\partial_q\log\Lambda$ is regular. The two gauges differ by $e^{\chi T_r}$ with $\chi=-(n-1)\arg z$, which is single valued on their simply connected overlap, so together they define one smooth connection. Reference~\cite{ST2009} evaluated this charge only for $n=2^k$; the restriction is unnecessary.

\section{Scope and implications}\label{sec:beyond}

\subsection{Which problem has no smooth real instantons?}

Theorem~\ref{thm:main} is a statement about the first-order action with a free boundary connection and the boundary problem \eqref{eq:critical}. It does not imply that non-flat self-dual connections on AdS$_4$ are absent; Sec.~\ref{sec:example} gives explicit examples. They fail to be saddles because no auxiliary field satisfies the remaining equations along with the source-free boundary equation.

Changing the variational problem changes the conclusion. If the boundary connection is held fixed, its variation vanishes and \eqref{eq:NBF} is not required. HRSS state that Dirichlet conditions are incompatible, for any $\theta$, with nontrivial self-dual configurations (Sec.~2.2 of Ref.~\cite{HRSS2026}). We read this as a statement about source-free Dirichlet data: tangential components $A_i$ on $S^3$ that vanish up to a gauge transformation. Then the tangential field strength $F_{jk}$ vanishes on $S^3$, self-duality gives, in the local axes of \eqref{eq:localdual}, $F_{1i}=\frac12\lc_{1ijk}F_{jk}=0$ for the normal components, and Proposition~\ref{prop:uc} gives $F=0$, in agreement with their statement. A prescribed non-flat boundary connection, by contrast, is a boundary source. With such data, a restricted BPST field together with $B=0$ solves the first-order equations; its size-changing deformation is not allowed because it alters the prescribed data. Similarly, if a source $J^\nu$ is added to the boundary equation, $n_\mu\big(B^{\mu\nu}+c(\star F)^{\mu\nu}\big)=J^\nu$, every smooth self-dual $A$ is a solution with $B=0$ and $J^\nu=c\,n_\mu F^{\mu\nu}$. These examples solve boundary problems with data chosen to admit the given field.

A nonzero Pontryagin coupling, equivalently $c\ne0$, is also essential. Without the Pontryagin term the boundary equation would read $n_\mu B^{\mu\nu}|_{S^3}=0$. Then $B=0$ completes every self-dual connection, and \eqref{eq:pair} puts no condition on $Q$. The theorem is therefore not a statement about every action that might be called a BF theory. Added boundary functionals, weaker fall-off at the boundary, or singular fields each call for their own variational and regularity analysis.

\subsection{Complex saddles and the path integral}\label{sec:pathint}

For complex connections, Theorem~\ref{thm:stationary} still holds, but the dilation argument then only requires the complex number $\int_{S^3}F^a_{\mu\nu}F^a_{\mu\nu}\,\dd\Omega_3$ to vanish, which does not force $F=0$. Complex saddles are therefore left open. Complex saddles do play a role in analytically continued gauge theories~\cite{WittenACS,TSH,HTS}, but whether they exist here and whether they lie on the relevant integration contour has to be shown for this problem.

On a self-dual connection, the formal weight is
\begin{equation}
 \exp(-\kappa Q[A]),\qquad \kappa=\frac{8\pi^2}{g_+^2}.
 \label{eq:Zlocal}
\end{equation}
The theorem excludes smooth, real, non-flat stationary points of this weight that can be completed by an auxiliary field. HRSS suggest that nonperturbative effects scale as powers of $e^{-1/g_+^2}$ (Sec.~3.1 of Ref.~\cite{HRSS2026}). If such effects are present, they do not come from smooth real saddles. The theorem does not exclude contributions from complex saddles or from noncompact ends of field space.

Along the restricted BPST family, $Q\to0$ as $d\to-\infty$ and $Q\to1$ as $d\to+\infty$. At the first end, the field inside the ball becomes weak. A small instanton with its center fixed inside the ball approaches the second end by concentrating its curvature at a point. No smooth non-flat saddle lies in between. A contribution from either end would require a choice of integration cycle, measure, and fluctuation determinant, together with control of the singular limit, and the classical argument provides none of these. In particular, the absence of smooth real instanton saddles does not mean that the theory is trivial beyond perturbation theory.

\section{Conclusions}\label{sec:conclusions}

In the first-order formulation of self-dual Yang-Mills theory on AdS$_4$, a classical saddle needs more than a self-dual connection: it needs an auxiliary field that satisfies a bulk equation and a boundary equation. The second-order boundary condition, which already excludes smooth real non-flat saddles at generic finite couplings, becomes vacuous in the self-dual limit. The first-order boundary equation does not. It forces the Pontryagin charge to be stationary under all self-dual deformations, and the dilation of the conformal ball is a deformation that changes the charge of every non-flat real field. When both fields are smooth up to the conformal boundary and the gauge field is real, the vacuum is the only saddle. This holds for every compact gauge group, without symmetry assumptions, and the auxiliary field vanishes as well.

The exact AdS$_4$ solutions remain useful self-dual configurations, but their charges must include the surface term at the conformal boundary. With that term, the charge of a restricted BPST instanton of any center and size is known in closed form. The inversion-symmetric half-charge value of 2009 is unchanged, and the higher-power solutions have charge $(n-1)/2$ for every $n\ge2$.

The remaining questions concern the existence and contribution of complex non-Abelian saddles and possible contributions from singular ends of field space. These require an analysis of the integration cycle and measure beyond the classical boundary problem studied here.

\appendix
\section{The reduced charge and the boundary term}\label{app:4d}\label{app:id}

We derive the reduction \eqref{eq:redaction} and the pointwise identity \eqref{eq:pointwise} used in the corrected charge \eqref{eq:Qcorrected}. Computing the reduction directly from the four-dimensional field strength fixes the normalization independently of the BPST identification.

\paragraph*{The reduced charge.}
We use the coordinates $(q,w,\vartheta,\varphi)$ and the moving basis $T_r,T_\vartheta,T_\varphi$ of the Lie algebra introduced before Eq.~\eqref{eq:ansatz}. Let $\mu$ take the values $q,w$, with $D_\mu\phi$ as in Sec.~\ref{sec:example}, and write $D_\mu\phi_1\equiv\mathrm{Re}\,D_\mu\phi$ and $D_\mu\phi_2\equiv\mathrm{Im}\,D_\mu\phi$. The independent components of the $SU(2)$ field strength are
\begin{align}
F_{qw}&=f_{qw}T_r, & f_{qw}&=\partial_q\mathfrak a_w-\partial_w\mathfrak a_q,
\nonumber\\
F_{\mu\vartheta}&=(D_\mu\phi_1)T_\vartheta+(D_\mu\phi_2)T_\varphi,
&F_{\mu\varphi}&=\sin\vartheta\big[(D_\mu\phi_1)T_\varphi-(D_\mu\phi_2)T_\vartheta\big],
\nonumber\\
F_{\vartheta\varphi}&=(|\phi|^2-1)\sin\vartheta\,T_r .
\label{eq:curvcomponents}
\end{align}
They follow from \eqref{eq:conventions}, the derivatives of the moving basis, and $[T_r,T_\vartheta]=T_\varphi$, $[T_\varphi,T_r]=T_\vartheta$, $[T_\vartheta,T_\varphi]=T_r$. With the flat metric $\dd s^2=\dd q^2+\dd w^2+q^2(\dd\vartheta^2+\sin^2\vartheta\,\dd\varphi^2)$,
\begin{equation}
\frac14F^a_{\mu\nu}F^{a\mu\nu}
=\frac12 f_{qw}^2+\frac{|D\phi|_0^2}{q^2}
 +\frac{(1-|\phi|^2)^2}{2q^4},
\qquad |D\phi|_0^2=|D_q\phi|^2+|D_w\phi|^2,
\label{eq:curvnorm}
\end{equation}
and the angular integral, with $\sqrt g=q^2\sin\vartheta$, gives
\begin{equation}
\frac14\int_{\Ball}\dd^4y\,F^a_{\mu\nu}F^{a\mu\nu}
=4\pi\int_{D_+}\dd q\,\dd w\left[
\frac{q^2}{2}f_{qw}^2+|D\phi|_0^2
+\frac{(1-|\phi|^2)^2}{2q^2}\right].
\label{eq:directred}
\end{equation}
For a self-dual field, $f_{qw}=-(1-|\phi|^2)/q^2$ by \eqref{eq:vortex}, and $\frac14\int F^a_{\mu\nu}F^{a\mu\nu}\,\dd^4y=8\pi^2Q$ by \eqref{eq:Q}. Substitution gives \eqref{eq:redaction}. It is the sum of the covariant-gradient and potential terms, not the Abelian flux alone, that reproduces the four-dimensional charge.

\paragraph*{The pointwise identity.}
Write $m^2=|\phi|^2$ and $\Delta_0=\partial_q^2+\partial_w^2=4\partial_u\partial_{\bar u}$, with $D_u=\frac12(D_q-iD_w)$ and $D_{\bar u}=\frac12(D_q+iD_w)$. Since $[D_q,D_w]\phi=if_{qw}\phi$, we have $[D_{\bar u},D_u]\phi=\frac12f_{qw}\phi$. The first vortex equation $D_{\bar u}\phi=0$ gives $\partial_{\bar u}m^2=\phi\,\overline{D_u\phi}$ and
\begin{equation*}
\partial_u\partial_{\bar u}m^2=|D_u\phi|^2
+\phi\,\overline{[D_{\bar u},D_u]\phi}
=|D_u\phi|^2+\tfrac12f_{qw}m^2 .
\end{equation*}
Again because $D_{\bar u}\phi=0$, $|D\phi|_0^2=2|D_u\phi|^2$, and the second vortex equation gives $f_{qw}=-(1-m^2)/q^2$. Multiplying by $2q^2$ and using $\Delta_\gamma=q^2\Delta_0$ and $|D\phi|^2_\gamma=q^2|D\phi|^2_0$,
\begin{equation*}
\tfrac12\Delta_\gamma m^2=|D\phi|^2_\gamma-m^2(1-m^2),
\end{equation*}
which is \eqref{eq:pointwise}, since $m^2(1-m^2)+(1-m^2)^2=1-m^2$.

\paragraph*{The boundary term.}
Inserting \eqref{eq:pointwise} into \eqref{eq:redaction}, the term $1-|\phi|^2$ yields $\Vf$, and the Laplacian term gives
\begin{equation}
Q-\Vf=\frac1{4\pi}\int_{D_+}\Delta_0|\phi|^2\,\dd q\,\dd w
=\frac1{4\pi}\oint_{\partial D_+}\partial_\perp|\phi|^2\,\dd s .
\label{eq:directboundary}
\end{equation}
For smooth fields, the axis $q=0$ does not contribute because $1-|\phi|^2=O(q^2)$ it is present; however, the semicircle $|u|=1$, the conformal boundary of AdS$_4$, generally does. This is the surface term missing from the 2009 flux calculation.

\section*{Acknowledgments}

I used two AI language models, OpenAI's ChatGPT and Anthropic's Claude,
to carry out and check analytic and numerical calculations, and to draft
and revise the manuscript. The boundary term omitted from the charge
calculation of Sar\i o\u glu and Tekin (2009)~\cite{ST2009} was first
identified during exchanges with these models; its derivation from the
four-dimensional curvature and its checks against the BPST family are
given in Sec.~\ref{sec:example} and the appendix. Working with these
models was a humbling experience.

\end{document}